\documentclass[fleqn,usenatbib]{mnras}
\usepackage[T1]{fontenc}
\usepackage{ae,aecompl}
\usepackage{graphicx}
\usepackage{amsmath}
\usepackage{amssymb}
\usepackage{txfonts}

\title[The $M_{\rm bh}$-$\sigma$ relation]
{Invoking the virial theorem to understand the impact of (dry) mergers on
  the $M_{\rm bh}$-$\sigma$ relation}

\author[Graham]
{
Alister W.\ Graham$^{1,2}$\thanks{E-mail: AGraham@swin.edu.au}
\\
$^1$ Centre for Astrophysics and Supercomputing, Swinburne University of
Technology, Hawthorn, VIC 3122, Australia
\\ 
$^2$ OzGrav-Swinburne, Centre for Astrophysics and Supercomputing, Swinburne
University of Technology, Hawthorn, VIC 3122, Australia
}

\date{Accepted XXX. Received YYY; in original form ZZZ}
\pubyear{2022}

\begin{document}
\label{firstpage}
\pagerange{\pageref{firstpage}--\pageref{lastpage}}
\maketitle

\begin{abstract}

While dry mergers can produce considerable scatter in the 
(black hole mass, $M_{\rm bh}$)--(spheroid stellar mass, $M_{\rm *,sph}$) 
and $M_{\rm bh}$--(spheroid half-light radius, $R_{\rm e,sph}$) diagrams, 
the virial theorem is used here to explain why the scatter about the $M_{\rm
  bh}$--(velocity dispersion, $\sigma$) relation remains low in the face of
such mergers.  Its small scatter has been claimed as evidence of feedback from
active galactic nuclei (AGNs).  However, it is shown that galaxy mergers 
also play a significant role.  The 
collision of two lenticular (S0) galaxies is expected to yield three types of
merger product (a core-S\'ersic S0, an ellicular ES,e or an elliptical E galaxy), depending on the
remnant's orbital angular momentum. 
It is shown that the major merger of two S0 galaxies with $M_{\rm
  *,sph}\sim10^{11}$~M$_\odot$ advances the system along a slope of $\sim$5 in
the $M_{\rm bh}$-$\sigma$ diagram, while a major E$+$E galaxy merger moves a system slightly along a
trajectory with a slope 
of $\sim$9.  Mergers of lower-mass S0 galaxies with $M_{\rm
  *,sph}\sim10^{10}$~M$_\odot$ move slightly along a trajectory with a slope of $\sim$3,
thereby further contributing to the steeper distribution for the E (and Es,e) galaxies
in the $M_{\rm bh}$-$\sigma$ diagram, reported here to have a slope of
7.27$\pm$0.91, compared to the S0 galaxies which have a slope of
5.68$\pm$0.60.  This result forms an important complement to the AGN feedback
models like that from Silk \& Rees, providing a more complete picture of galaxy/(black
hole) coevolution. 
It also has important implications for nanohertz gravitational wave research.

\end{abstract}


\section{Introduction}

Before it was observed, \citet{1998A&A...331L...1S} 
predicted a relation between a halo's central
supermassive black hole mass, $M_{\rm bh}$, and the host halo's velocity
dispersion\footnote{This $\sigma$ is the assumed constant velocity dispersion
  of a dark matter halo described by an isothermal sphere.}, $\sigma$, such
that $M_{\rm bh} \propto \sigma^5$.  \citet{1998MNRAS.300..817H} built on this
by predicting $M_{\rm bh} \propto M_{\rm halo}^{1.67}$.  Using rotational
velocities to probe the halo mass, \citet{2002ApJ...578...90F} reported
$M_{\rm bh} \propto M_{\rm halo}^{1.65}$-$M_{\rm halo}^{1.82}$, where $M_{\rm
  halo}$ is the total (dark matter, stellar, and gas) mass.  Although, using
rotational velocities for 48 spiral (S) galaxies, \citet{2019ApJ...877...64D}
have since measured the notably steeper relation\footnote{Coupled with $M_{\rm
    bh}\propto M_{\rm *,gal}^{3.05}$ \citep{2018ApJ...869..113D}, this implies
  $M_{\rm halo}/M_{\rm *,gal} \propto M_{\rm *,gal}^{-0.3}$, that is, lower
  mass spiral galaxies have higher halo-to-stellar mass ratios.}  $M_{\rm bh}
\propto M_{\rm halo}^{4.35\pm0.66}$.  Curiously, working with the stellar
masses of bulges/spheroids, $M_{\rm *,sph}$, \citet{Graham:Sahu:22a} have
reported $M_{\rm bh} \propto M_{\rm *,sph}^{1.64\pm0.17}$ and $M_{\rm bh}
\propto M_{\rm *,sph}^{1.53\pm0.15}$, respectively, for elliptical (E)
galaxies and the bulges of lenticular (S0) galaxies.  The non-linear nature of
the low- and high-mass end of the $M_{\rm bh}$-$M_{\rm *,sph}$ relation(s)
was, however, already noted by \citet[][and references
  therein]{2016ASSL..418..263G} and \cite{2019ApJ...876..155S}, respectively.

%

Many potential pathways exist for massive black holes to grow at the centres
of galaxies \citep[e.g.,][]{1984ARA&A..22..471R}.  One of these is engendered
through `dry' galaxy mergers rather than active galactic nuclei (AGNs), leading
to the generation of long-wavelength gravitational waves when the black holes
merge \citep[e.g.,][]{1974bhgw.book.....R, 1980Natur.287..307B,
  2011ApJ...732...89K, 2013CQGra..30x4009S} and the likely creation of E
galaxies.  Indeed, \citet{2019ApJ...876..155S} have revealed that E galaxies
and their massive black holes represent a subsequent generation in the family
tree of galaxy-(black hole) coevolution.  In the $M_{\rm bh}$-$M_{\rm *,sph}$
diagram, E galaxies are offset from the bulges of S0 galaxies by almost an
order of magnitude in the $M_{\rm bh}$ direction.  This is not simply due to
the exclusion of the S0 galaxies'disc mass because they are additionaly offset
by a factor of $\sim$2 in the $M_{\rm bh}$-$M_{\rm *,gal}$ diagram.
\citet{Graham:Sahu:22a} explain how this offset reflects the creation of E
galaxies from the merger of S0 galaxies, which themselves follow a quadratic
$M_{\rm bh}$-$M_{\rm *,gal}$ relation.  That is, AGN feedback
\citep[e.g.][]{2010ApJ...725..556S, 2014MNRAS.444.2355C, 2014ARA&A..52..589H}  has not 
established the (black hole)-galaxy relation for the E galaxies, but rather the
presence of a central heat source --- dubbed a `Benson Burner' after
\citet{2003ApJ...599...38B} --- has maintained the merger-established
$M_{\rm bh}$-$M_{\rm *,sph}$ (and $M_{\rm bh}$-$M_{\rm *,gal}$) connection
by preventing gas from cooling to form new stars or reignite the central
quasar. Within clusters, ram-pressure stripping from the hot X-ray gas
likely performs a similar role for the more centrally-located galaxies
\citep[e.g.][]{2006ApJ...647L..21H, 2015A&A...574A.126G}. 

This order of magnitude offset in the $M_{\rm bh}$-$M_{\rm *,sph}$ diagram is
mirrored in the $M_{\rm bh}$-$R_{\rm e,sph}$ diagram, providing an
opportunity to use the spheroids' effective half-light radii, $R_{\rm e,sph}$, in the virial theorem
\citep[$M_{\rm dyn,sph} \propto \sigma^2 R_{\rm e,sph}$: e.g.,][]{2021isd..book.....C} 
to give some insight into 
the distribution in the $M_{\rm bh}$-$\sigma$ diagram
\citep{2000ASPC..197..221M, 2000ApJ...539L...9F, 2000ApJ...539L..13G}.  Here,
$\sigma$ is the stellar velocity dispersion of the stars. The 
low scatter about the $M_{\rm bh}$-$\sigma$ relation has long been heralded
as a sign that AGN feedback regulates the coevolution of galaxies and their
central black holes.  However, for the E galaxies, it is not AGN feedback but
mergers which have established their $M_{\rm bh}/M_{\rm *,sph}$ ratio. 
In this paper, it is 
explained how mergers also maintain a tight $M_{\rm bh}$-$\sigma$ relation,
with small departures expected to cause a steepening of the relation for 
the E (and ES) galaxies. 

\begin{figure*}
\begin{center}
\includegraphics[trim=0.0cm 0cm 0.0cm 0cm, width=1.0\textwidth, angle=0]{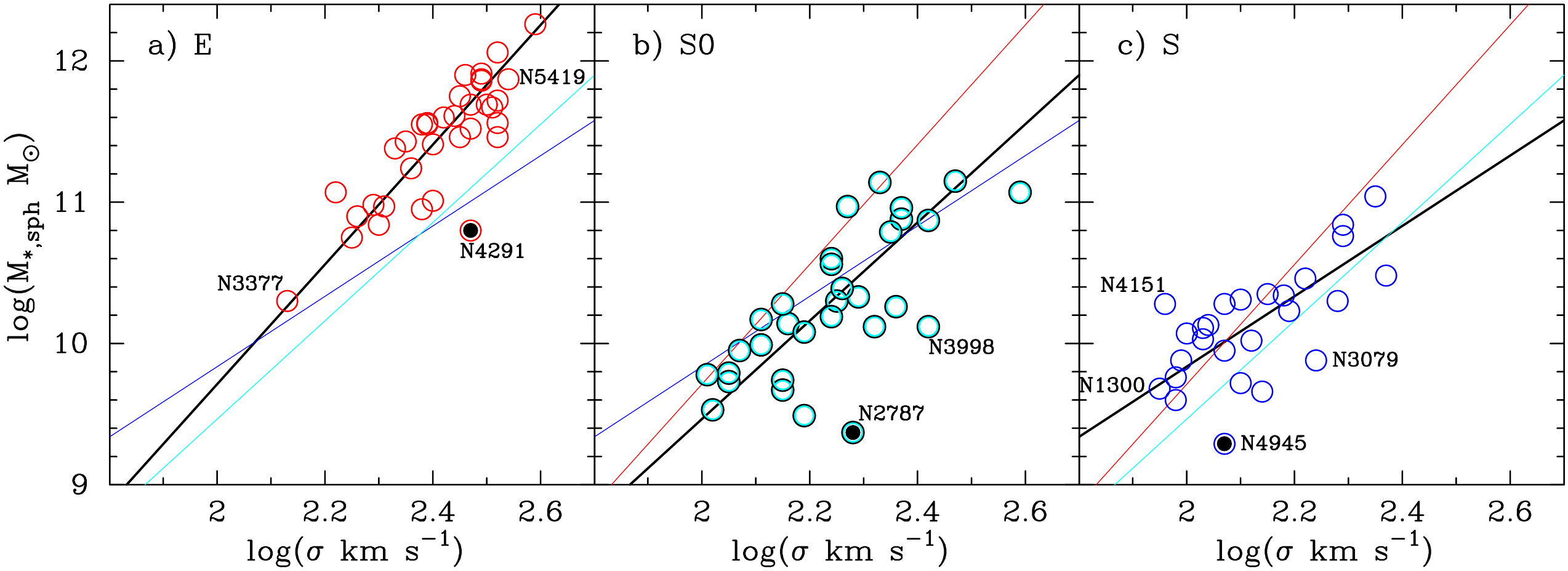}
\caption{Spheroid (not galaxy) stellar mass plotted against central velocity
  dispersion for 92 systems. 
The three galaxies over-plotted with a black circle are excluded from the
regression analyses, while the other labelled galaxies are retained. 
The slopes are 4.24$\pm$0.54, 3.48$\pm$0.47, and 
2.49$\pm$0.56 for the E (= E and ES,e), S0 (= S0 and ES,b) and S samples, respectively.
For reference, when using $M_{\rm *,tot}$ rather than $M_{\rm *,sph}$, the
slopes are 4.14$\pm$0.55, 2.71$\pm$0.45, and 1.90$\pm$0.35. 
%
}
\label{Fig-3-1}
\end{center}
\end{figure*}

Section~\ref{Sec_data} describes the data sample, shown in
Sections~\ref{Sec_data} and \ref{Sec_AaD}. In essence, it is used to reveal
why the large, merger-induced 
jumps in $M_{\rm *,sph}$ and $R_{\rm e,sph}$ --- once the disc stars from
S0 galaxies become the outer regions of a merger-built E galaxy --- 
leave little room for increases in $\sigma$. 
It is shown  that such collisions roughly move galaxies along the $M_{\rm
  bh}$-$\sigma$ relation or produce small offsets leading to a 
steepening at mid-to-high masses. 
A brief summary is provided in Section~\ref{Sec_Summary}. 

In Appendix~\ref{Sec_Apdx}, 
a range of dynamical-to-stellar mass ratios are presented, 
along with caveats about inferring the presence of 
dark matter in E galaxies from these ratios. 
Appendix~\ref{App_Out} reports on three galaxies flagged for exclusion in
Section~\ref{Sec_Out}.

\section{The Data Sample}\label{Sec_data}

\subsection{Data Source}\label{Sec_DS} 

Correcting a false assumption in \citet{2016MNRAS.460.3119S} regarding 
the use of, what turned out to be inconsistent, stellar mass-to-light ratios, 
the sample of galaxies with directly measured black 
hole masses is not biased with respect to the galaxy
population at large \citep{SahuGrahamHon22}.

This research forms an extension of \citet{Graham:Sahu:22a} and
\citet{Graham:Sahu:22b}, which was based on an initial sample of 104 galaxies with
directly measured supermassive black hole masses and Spitzer Space Telescope (SST)
imaging at 3.6~$\mu$m.  Two of those galaxies were
bulgeless spiral galaxies (NGC~4395 and NGC~6926), 
and an additional nine were excluded from the regression analysis
due to the reasons given in \citet[][see the end of their Section~2.1]{Graham:Sahu:22a}.
Briefly, five of the nine excluded galaxies are well-recognised merger
remnants.  Shown here in the
figures, they are explained in \citet{Graham:Sahu:22b} in terms of (cold gas)-rich
mergers in which the system's angular momentum is not cancelled and thus a disc is
present in the merger remnant. 
These galaxies are not included in the regression analyses. 
The four other galaxies are  
Circinus \citep[another unrelaxed spiral galaxy][]{2012MNRAS.425.1934F}\footnote{It is speculated here that
  the apparent bulge component of this galaxy \citep{2019ApJ...873...85D} 
might be the inner part of an anti-truncated disc, blurring the definition of 
what is a bulge.}, 
NGC~5055 (a spiral galaxy assigned an incorrect black hole mass)\footnote{The
  often reported mass is the total mass interior to 300~pc \citep{2004A&A...420..147B}.}, 
NGC~4342 (a stripped S0 galaxy with a questionable spheroid mass), and 
the S0 galaxy NGC~404 \citep{2017ApJ...836..237N}, which is 
the only galaxy in the sample with $M_{\rm bh} < 10^6$ M$_{\odot}$
and therefore potentially carrying too much weight in regression analyses.  
While such low-mass black holes are much sought after, there are not many
because it is challenging to 
spatially-resolve their sphere of gravitational influence
\citep{2013degn.book.....M} --- although future instruments hold much promise
\citep[e.g.,][their Section~5.1]{2021ApJ...923..246G}.  Moreover, the velocity
dispersion can be influenced, if not dominated, by the disc and, when present, the bar in these 
low-mass galaxies.  This latter issue is left for future work given this paper's focus on
roughly the upper-half of the $M_{\rm bh}$-$\sigma$ diagram, where E galaxies
reside.  
Here, NGC~7457 is also excluded.  As discussed in \citet{2019ApJ...887...10S}, 
this peculiar S0 galaxy was already recognised as having an oddly low velocity dispersion
relative to its peers, plus other unusual kinematic properties such as
cylindrical rotation about its major axis 
\citep{2002ApJ...577..668S, 2019MNRAS.488.1012M}.

The above exclusions collectively reduced the working sample to 92, involving 
35 E galaxies (including ten ES,e galaxies), 31 S0 galaxies (including two
core-S\'ersic S0 galaxies and four ES,b galaxies), and 26 S
galaxies.\footnote{\citet{1966ApJ...146...28L} introduced the ES galaxy
  notation, with the subtypes ES,e and ES,b introduced in \citet{Graham:Sahu:22b}.}  Although,
as seen in the following subsection, each of these three subsamples is further
reduced by one.  For the regression, following \citet{Graham:Sahu:22a}, the
ES,e and ES,b galaxies \citep[][see their Table~1]{Graham:Sahu:22b} are 
treated as though they are E and S0 galaxies, respectively.  Galaxies excluded
from the regression analyses are still plotted in Section~\ref{Sec_AaD}.

The spheroid sizes have come from \citet[][their Table~1]{Graham:Sahu:22a} 
and are based on the updated distances reported there.  These sizes were derived from the
multicomponent galaxy decompositions that are shown for every galaxy in \citet{2016ApJS..222...10S},
\citet{2019ApJ...873...85D}, \citet{2019ApJ...876..155S}, or the Appendix of
\citet{Graham:Sahu:22a}. 
The spheroid stellar masses stem from the spheroid magnitudes
obtained in the above galaxy decompositions, coupled with 
a colour-dependent stellar mass-to-light ratio.  These ratios are based on 
the \citet{2013MNRAS.430.2715I} stellar population 
models, as adjusted by Equation~4 in \citet{Graham:Sahu:22a} to a
\citet{2002Sci...295...82K} initial mass function. 
 
With one exception (NGC~1300), the central stellar velocity dispersions,
$\sigma$ --- tabulated in \citet{2019ApJ...887...10S} and taken from the
HyperLeda database\footnote{http://leda.univ-lyon1.fr}
\citep{2003A&A...412...45P} --- are used.  NGC~1300 was previously flagged by
\citet{2019ApJ...887...10S} as a notable outlier in the $M_{\rm
  *,sph}$-$\sigma$ diagram.  Their velocity dispersion for NGC~1300 appears
to be in error.\footnote{HyperLeda reports values from 145 to 304 km s$^{-1}$
  taken from two studies.}$^,$\footnote{It is noted that NGC~1300 was also
  flagged by \citet[][their Fig.~8]{Graham:Sahu:22a} as something of an
  outlier in the $M_{\rm bh}$-$M_{\rm *,sph}$ diagram. This may in part be due
  to a possibly missed barlens (dominant over $\sim$10--35$\arcsec$) 
  in the decomposition \citep{2019ApJ...873...85D}, but 
  may also reflect an overly large black hole mass measurement \citep{2005MNRAS.359..504A}.} \citet[][their
  Table~1]{2005ApJS..160...76B} report values around 82 to 90 km s$^{-1}$.  A
value of 90$\pm$12 km s$^{-1}$ is adopted here.


\subsection{Checking for outliers in the $M_{\rm *,sph}$-$\sigma$ diagram}
\label{Sec_Out}

Before proceeding, the $M_{\rm *,sph}$-$\sigma$ diagram
(Fig.~\ref{Fig-3-1}) was inspected for potentially biasing
data points.  Three such galaxies were detected 
(NGC~2787, NGC~4291; and NGC~4945). 
Failing to identify and remove outlying data, not representative of the population at
large, can bias a regression.  This may occur when the outlying data point is
located near the end of a trend, causing a fitted regression line to shift 
towards the outlier. 
Sometimes the reason for discrepant data is
identified and can be corrected, while other times, it remains unknown.  
These three galaxies identified in Fig.~\ref{Fig-3-1} are discussed in Appendix~\ref{App_Out}.  They are
excluded from the regression analyses that follow and together with
NGC~7457, mentioned in the previous subsection, 
results in a sample that is smaller by four than used in \citet{Graham:Sahu:22a}.

\subsection{Galaxies built from dry mergers} 

\begin{figure}
\begin{center}
\includegraphics[trim=0.0cm 0cm 0.0cm 0cm, width=1.0\columnwidth, angle=0]{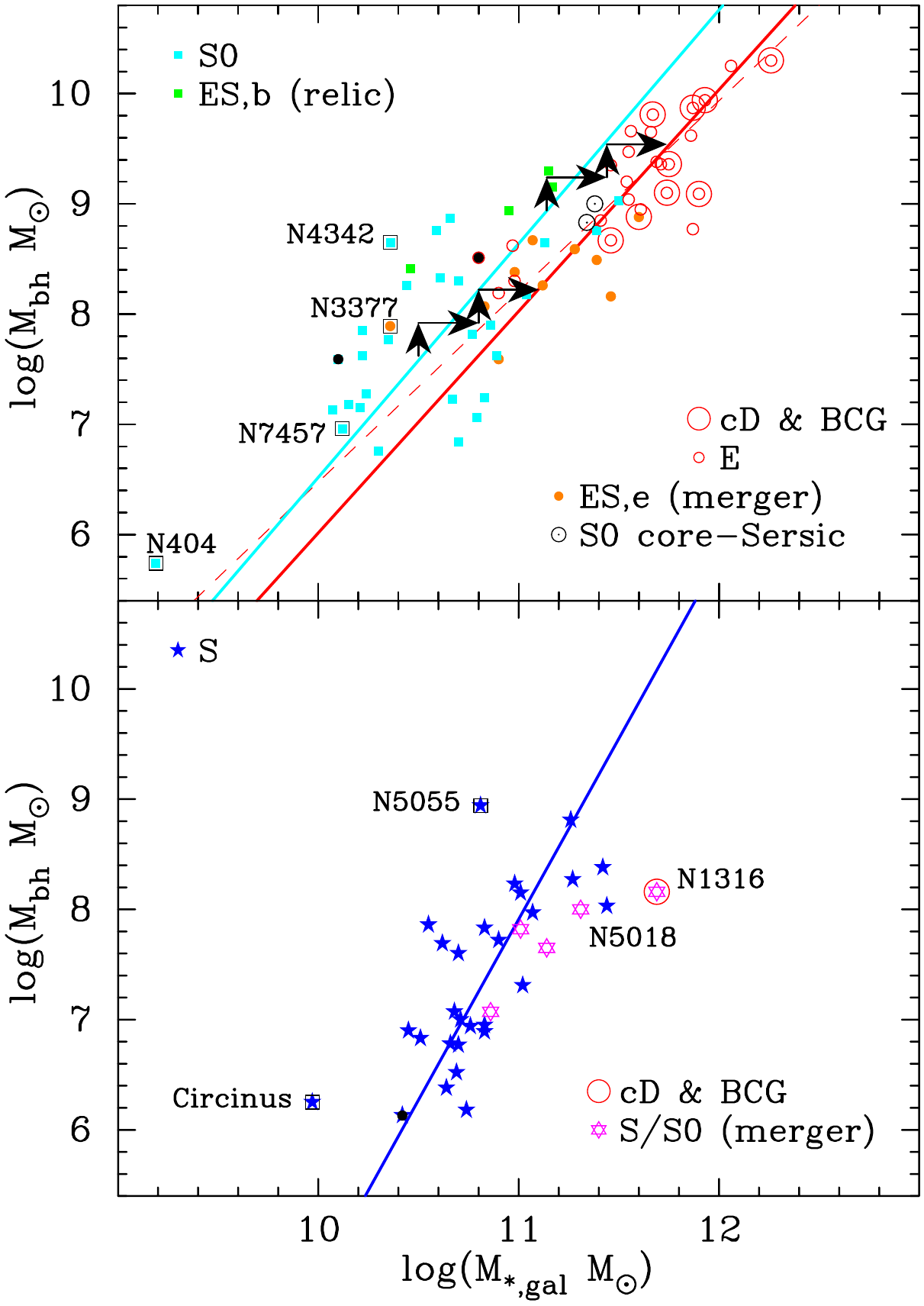}
\caption{Modification of Fig.~5 from \citet{Graham:Sahu:22a}.
Black hole mass versus galaxy stellar mass. 
Six labelled points enclosed in a black square (the three S0, two S, and one ES,e galaxies mentioned in
Section~\ref{Sec_DS}), plus three points over-plotted with a black 
circle (NGC~2787, NGC~4291, and NGC4945 from Figure~\ref{Fig-3-1}), 
are excluded from the fits, as are the five S/S0 mergers (pink hexagon). 
For 33 E$+$ES,e galaxies (excluding NGC~3377) plus two core-S\'ersic S0
galaxies (perhaps also built by somewhat dry mergers), one has: 
$\log(M_{\rm bh}/M_\odot) = (2.01\pm0.21)[ \log(M_{\rm bh}/M_\odot)
  -11.49]+(9.01\pm0.11)$ (red line).  Including NGC~3377 gives the dashed line
with a slope of 1.73$\pm$0.18. 
For 28 (S\'ersic S0)$+$ES,b galaxies: 
$\log(M_{\rm bh}/M_\odot) = (2.12\pm0.34)[ \log(M_{\rm bh}/M_\odot)
  -10.68]+(7.96\pm0.17)$ (cyan line). 
For 25 S galaxies: 
$\log(M_{\rm bh}/M_\odot) = (3.28\pm0.66)[ \log(M_{\rm bh}/M_\odot)
  -10.85]+(7.40\pm0.16)$ (blue line).
The four pairs of arrows denote equal-mass dry mergers, in which $M_{\rm bh}$
and $M_{\rm *,gal}$ double. 
}
\label{Fig-3-2}
\end{center}
\end{figure}

\citet[][their Fig.~8]{2019ApJ...876..155S} revealed that E galaxies are
offset from S0 galaxies in the $M_{\rm bh}$-$M_{\rm *,gal}$ diagram.
Fig.~\ref{Fig-3-2} helps explain this offset between the current sample of E
(and ES,e) and S0 (and ES,b) galaxies in the $M_{\rm bh}$-$M_{\rm *,gal}$
diagram.  The ES,e ellicular galaxies have previously been referred to as
`disc ellipticals' \citep{1988A&A...195L...1N, 1995A&A...293...20S}.
\citet{Graham:Sahu:22b} discusses why they likely formed from a major merger
event making them more akin to elliptical galaxies, while the ES,b ellicular
galaxies, also known as `compact galaxies' \citep{1968cgcg.bookR....Z,
  1971cscg.book.....Z} and 'relic red nuggets' or `relic galaxies'
\citep[e.g.,][]{2017MNRAS.467.1929F}, likely never accreted a large-scale
disc, making them more akin to classical bulges.  
Given that S\'ersic S0 galaxies have grown their discs through 
(angular momentum)-building (cold gas)-rich accretion and merger events
\citep[e.g.,][]{2015ApJ...804...32G, 2022MNRAS.514.3410H,
  2022MNRAS.511..607J}, while core-S\'ersic S0 galaxies 
are likely built from significant, relatively dry mergers, the two 
core-S\'ersic S0 galaxies are best shifted from the S0 to the E galaxy bin
when dealing with galaxy stellar masses.  Such a redesignation only affects
Fig.~\ref{Fig-3-2}, which is not used for our upcoming analysis of the $M_{\rm
  bh}$-$\sigma$ diagram.

Due to the (steeper than linear) 
$M_{\rm bh}\propto M_{\rm *,gal}^2$ relation for S0 galaxies, the (dry merger)-built
E galaxies will not move along this relation but effectively shift to the
right.  Additional major dry mergers will move systems along a path with a
slope of 1 in the $M_{\rm bh}$-$M_{\rm *,gal}$ diagram. 
It appears that (cold gas)-rich mergers involving S galaxies also produce merger
remnants shifted rightward of the $M_{\rm bh}\propto M_{\rm *,gal}^3$ relation for S galaxies. 

The \citep{1966ApJ...146...28L}
ES galaxies with intermediate-scale discs 
have previously been recognised, on a morphological basis, as an intermediary
between E and S0 galaxies \citep[e.g.,][and references
  therein]{2019MNRAS.487.4995G}.  With recourse to their $M_{\rm bh}/M_{\rm
  *,gal}$ ratios, Fig.~\ref{Fig-3-2} now shows their evolutionary connection.
The ES,e galaxies appear as something of a bridging population, an
intermediary step facilitating the speciation of E galaxies from S0 galaxies.
Through the mating of two S0 galaxies to produce an ES,e galaxy, and the
mating of two ES,e galaxies (or E$+$ES,e) to produce an E
galaxy, one can start to see the lineage and transformation of these
galaxies. Similarly, two S0 galaxies may merge to form an E galaxy, with the
subsequent merger of two E galaxies forming a bigger E galaxy, such as a
brightest cluster galaxy (BCG).  Mergers are a well-known phenomenon, but an
exploration of galaxy pairings, as systems evolve to higher masses in the $M_{\rm
  bh}$-$M_{\rm *,gal}$ diagram, has only recently been observed \citet[][their
  Figures~6--8]{Graham:Sahu:22a}.  
In the following section, this is extended to the observed $M_{\rm
  bh}$-$M_{\rm dyn,sph}$ and $M_{\rm bh}$-$\sigma$ diagrams.

\section{Analysis and Discussion}
\label{Sec_AaD} 

\subsection{Dynamical masses and black hole scaling relations}

\begin{figure}
\begin{center}
\includegraphics[trim=0.0cm 0cm 0.0cm 0cm, width=1.0\columnwidth, angle=0]{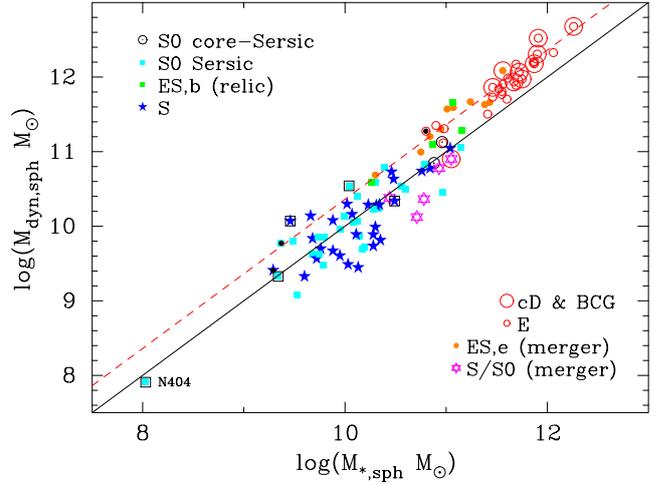}
\caption{$M_{\rm dyn,sph}$ ($\equiv 5\sigma^2 R_{\rm e,sph}$) plotted against $M_{\rm
    *,sph}$ for 102 ($=104-2$) spheroids.  Roughly, for the (merger-built)
  elliptical galaxies, $M_{\rm dyn,sph}/M_{\rm *,sph} = 2.3$, as shown by the
  dashed red line, while $M_{\rm dyn,sph}/M_{\rm *,sph} = 1$ for the bulges of
  the disc galaxies (solid black line).  As explained in the text, it would be premature to
  interpret these results as an increased dark matter fraction within
  the E galaxies relative to the bulges of disc galaxies.  The five systems
  enclosed in a black square, and the three systems over-plotted with a black
  circle, are the exclusions mentioned in the first and second part of
  Section~\ref{Sec_data}, respectively. They are not used in the upcoming
  regression analyses, nor are the five S/S0 mergers.} 
\label{Fig-3-3}
\end{center}
\end{figure}

Here, the dynamical mass is defined as 
\begin{equation} 
M_{\rm dyn} = 5 \sigma^2 R_{\rm e} / G, 
\label{Eq_dyn}
\end{equation}
with the virial coefficient of 5 taken from \citet{2006MNRAS.366.1126C}.  As
noted earlier, the equivalent-axis\footnote{The `equivalent axis', also known
  as the geometric-mean axis, is such that the radius $R_{\rm equiv} \equiv
  \sqrt{R_{\rm major}R_{\rm minor}}$ provides a light profile equivalent to
  that coming from a circularised version of the projected spheroid light.}
effective half-light radii of the spheroids, $R_{\rm e}$, are listed in
\citet{Graham:Sahu:22a} and the central velocity dispersions, $\sigma$, are
listed in \citet{2019ApJ...887...10S}.

Fig.~\ref{Fig-3-3} shows the spheroid dynamical mass (Equation~\ref{Eq_dyn})
plotted against the spheroid stellar mass, taken from \citet{Graham:Sahu:22a}.
On average, the E galaxies roughly have $M_{\rm dyn,sph}/M_{\rm
  *,sph}=2.3$, while the bulges of the disc galaxies have a lower ratio of around
1.  As discussed in Appendix~\ref{Sec_Apdx}, it is premature to conclude that
the higher ratio in the merger-built E galaxies is due to dark matter halos of
the progenitor disc galaxies contributing within (the now larger) 
$R_{\rm e,sph}$ of the E galaxies.
  
Using $M_{\rm dyn,sph} \approx  2.3 M_{\rm *,sph} \approx 5 \sigma^2 R_{\rm e,sph} /
G$, the E galaxies follow 
\begin{equation}
M_{\rm *,sph} \approx (5/2.3) \, \sigma^2 \, R_{\rm e,sph}/G.
\label{Eq_stellar}
\end{equation}

\begin{figure*}
\begin{center}
\includegraphics[trim=0.0cm 0cm 0.0cm 0cm, width=1.0\textwidth, angle=0]{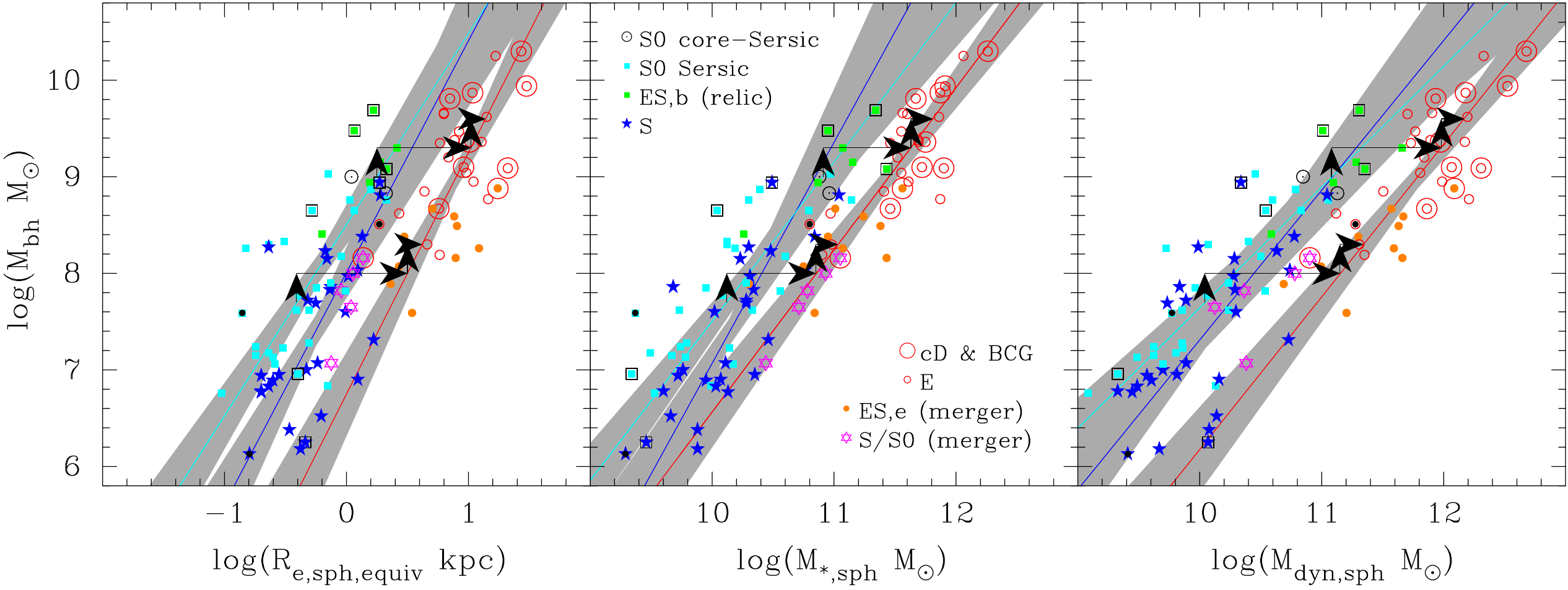}
\caption{Expanding Fig.~2 from \citet{Graham:Sahu:22b}, a 
  third panel has been added to show a proxy for dynamical mass ($\equiv 5\sigma^2
  R_{\rm e,sph}$).  Included are the regression lines and 1-sigma
  uncertainties (grey bands) for the late-type  galaxies, the
  elliptical galaxies (E and ES,e), and the bulges of relaxed
  early-type galaxies with discs (specifically, the S0 galaxies, including the two
  core-S\'ersic S0 galaxies and the ES,b galaxies).  These lines are 
  shown in blue, red, and cyan, respectively, and they are summarised in Table~1.  
  Plotted but not included in the Bayesian 
  analyses are five S/S0 mergers (pink hexagons), five (3 S0 $+$ 2 S) exclusions mentioned in
  Section~\ref{Sec_DS} (enclosed with a black square), and three outliers
  observed in Figure~\ref{Fig-3-1} (over-plotted with a black circle). 
  Three ES,b galaxies from \citet{Graham:Sahu:22b} (Mrk~1216, NGC~1271 and
  NGC~1277) without a Spitzer image analysis are also included in the figure
  but not the regression.  These green points re enclosed in a black square. 
  The arrows show shifts for dry equal-mass mergers, specifically 
   (i) a binary S0 galaxy merger initially with $M_{\rm
    bh}=0.5\times10^8$~M$_\odot$ in each galaxy, (ii) a binary E galaxy merger initially with
  $M_{\rm bh}=10^8$~M$_\odot$ in each galaxy, (iii) a binary S0 galaxy merger initially with $M_{\rm
    bh}=10^9$~M$_\odot$ in each galaxy, and (iv) a binary E galaxy merger initially with
  $M_{\rm bh}=2\times 10^9$~M$_\odot$ in each galaxy.  In each instance 
  $M_{\rm bh}$ doubles. 
}
\label{Fig-3-4}
\end{center}
\end{figure*}

Fig.~\ref{Fig-3-4} presents the $M_{\rm bh}$-$M_{\rm *,sph}$ and $M_{\rm
  bh}$-$R_{\rm e,sph}$ diagrams from \citet{Graham:Sahu:22a}, showing the
offset between the E galaxies and the bulges of the S0 galaxies.  An
additional third panel showing the dynamical mass of these spheroids, as per
Equation~\ref{Eq_dyn}, has been added.  Following \citet{Graham:Sahu:22a},
Bayesian model fitting using {\sc Stan} \citep{2017JSS....76....1C,
  Rstan:2016}\footnote{\url{https://mc-stan.org/}} has been employed.  The
approach is described in \citet[][their Appendix~A]{2019ApJ...873...85D} and
was used to obtain a symmetrical linear regression between the plotted
quantities.  The regressions are based on almost the same sample of 35 E and
32 ES/S0 galaxies shown in \citet[][their Table~1]{Graham:Sahu:22a}. As
discussed in Section~\ref{Sec_data}, the E sample is reduced by one (NGC~4291)
and the ES/S0 sample by two (NGC~2787 and NGC~7457).  Another distinction from
the figures in \citet{Graham:Sahu:22a} is that Fig.~\ref{Fig-3-4} provides a more detailed
morphological description using the designations given in 
\citet{Graham:Sahu:22b}.  This encompasses labelling the brightest cluster
galaxies (BCGs) and differentiating between two potential types of ES galaxy,
namely the compact ES,b systems and the larger ES,e systems, likely
been built by mergers of S0 galaxies. 

\begin{figure}
\begin{center}
\includegraphics[trim=0.0cm 0cm 0.0cm 0cm, width=1.0\columnwidth,
  angle=0]{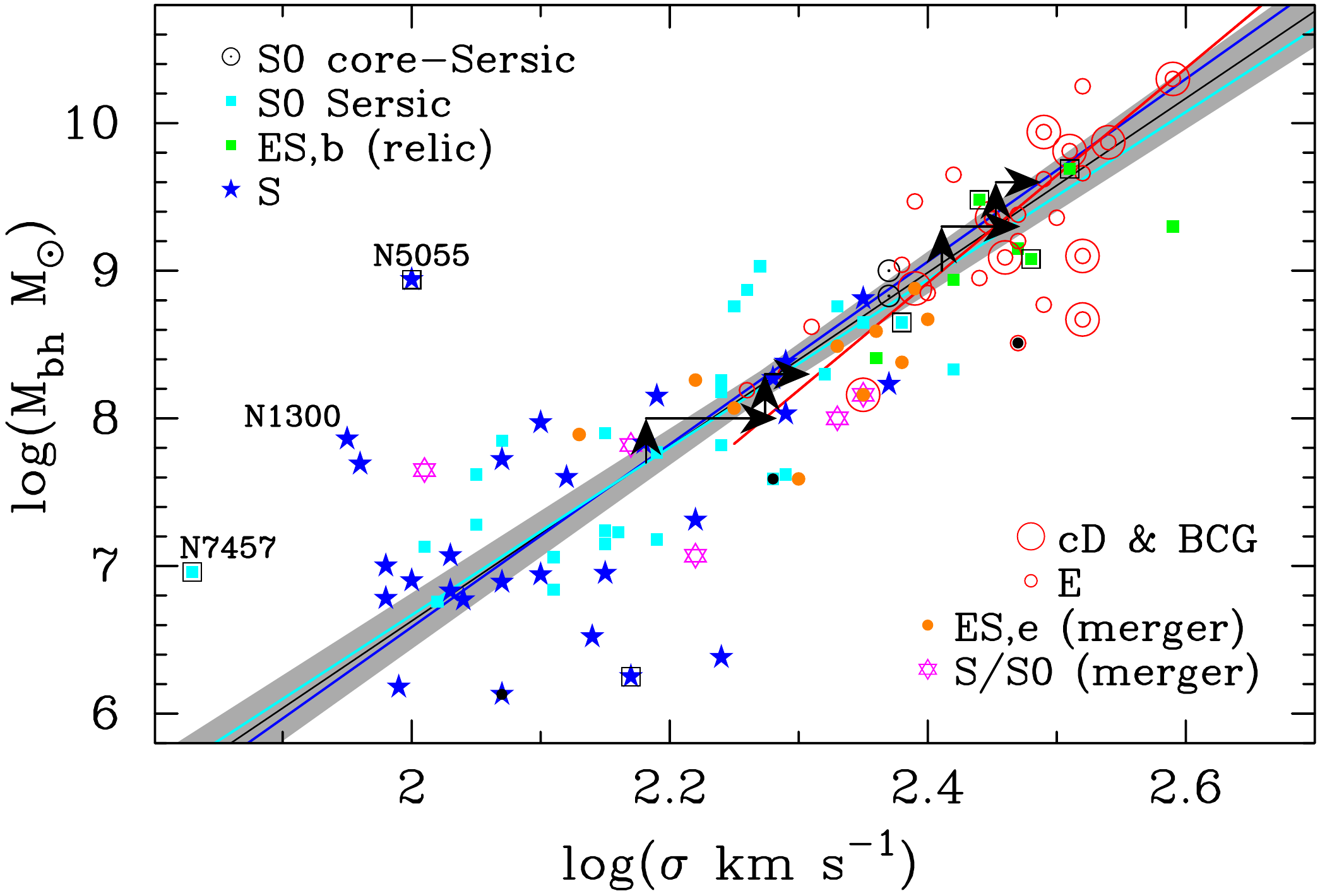}
\caption{$M_{\rm bh}$-$\sigma$ diagram similar to those in
  \citet{2019ApJ...887...10S} but now showing more of the galaxies' morphological
  detail and using the reduced sample with Spitzer imaging
  shown in Fig.~\ref{Fig-3-4}.  
Relations for the three different morphological types (E, S0, S) are presented in
Table~\ref{Table-IP13}.  The symbols, colours, and lines have the same meaning
  as in Figures~\ref{Fig-3-2} to \ref{Fig-3-4}, except now only one 1-sigma confidence band (gray
  shading) is shown for the black line, which represents all three galaxy types combined. 
  The vertical arrows show a doubling in black hole mass while the associated
  horizontal arrows are now the {\it predicted}, rather than measured, shifts in $\sigma$. 
  As explained in the text, they are based on the virial theorem and the observed shifts in
  Fig.~\ref{Fig-3-4} 
  associated with (i) a binary S0 galaxy merger initially with $M_{\rm
    bh}=0.5\times10^8$~M$_\odot$ in each galaxy, (ii) a binary E galaxy merger initially with
  $M_{\rm bh}=10^8$~M$_\odot$ in each galaxy, (iii) a binary S0 galaxy merger initially with $M_{\rm
    bh}=10^9$~M$_\odot$ in each galaxy, and (iv) a binary E galaxy merger initially with
  $M_{\rm bh}=2\times 10^9$~M$_\odot$ in each galaxy.  
}
\label{Fig-3-5}
\end{center}
\end{figure}

Knowing how $M_{\rm bh}$ varies with $R_{\rm e,sph}$ (Fig.~\ref{Fig-3-4},
left-hand side), one can determine from Equation~\ref{Eq_stellar} how $M_{\rm
  bh}$ must vary with $\sigma$ to match how $M_{\rm bh}$ varies with
$M_{\rm *,sph}$ (Fig.~\ref{Fig-3-4}, middle panel).
Although one could have bypassed the dynamical mass and simply inspected how
$M_{\rm bh}$ varies with $\sigma$ (Fig.~\ref{Fig-3-5}), doing so would not
have explicitly revealed the constraint from the virial theorem.  That is,
insight into why $M_{\rm bh}$ varies with $\sigma$ in the way it does
would have been absent.
The $M_{\rm bh}$-$\sigma$ relations in Fig.~\ref{Fig-3-5} shall be visited 
in the following subsection.  A couple of relevant observations
should, however, first be made. 

\begin{figure}
\begin{center}
\includegraphics[trim=0.0cm 0cm 0.0cm 0cm, width=1.0\columnwidth, angle=0]{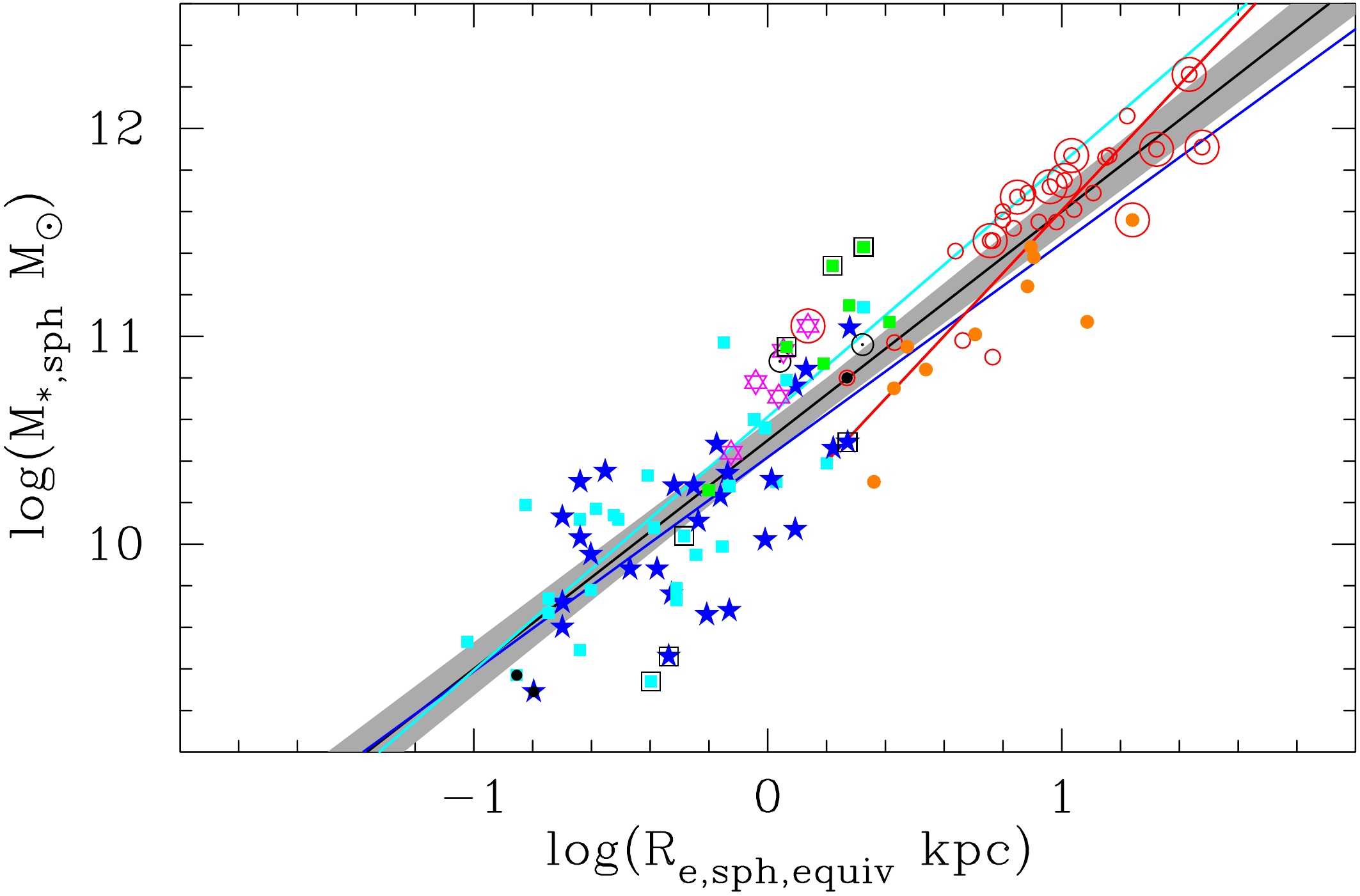}
\caption{Spheroid (stellar mass)-size diagram similar to Fig.~7 in
  \citet{Graham:Sahu:22a} but showing more detail regarding the host galaxy's
  morphological type.  The symbols,
  colours, lines, and shading have the same meaning as in Figure~\ref{Fig-3-5}. 
The relations for the three different 
  morphological types, and the combined sample, are reported in Table~\ref{Table-IP13}, 
%
%
}
\label{Fig-3-6}
\end{center}
\end{figure}


\begin{table}
\centering
\caption{Black hole mass scaling relations}\label{Table-IP13}
\begin{tabular}{lcccc}
\hline
Type   &   slope (A) & mid-point (C) & intercept (B) & $\Delta_{\rm rms}$ \\
\hline
\multicolumn{5}{c}{$\log(M_{\rm bh}/{\rm M}_\odot) = A[\log(R_{\rm e,sph,eq}/{\rm kpc})-C] +B$} (Fig.~\ref{Fig-3-4}b)\\
E   (34)     &  2.50$\pm$0.35  &     0.90  & 9.00$\pm$0.12 &  0.58 \\
S0  (30)     &  1.98$\pm$0.25  &  $-$0.23  & 8.05$\pm$0.15 &  0.52 \\
S   (25)     &  2.40$\pm$0.42  &  $-$0.25  & 7.41$\pm$0.16 &  0.67 \\
\hline
\multicolumn{5}{c}{$\log(M_{\rm bh}/{\rm M}_\odot) = A[\log(M_{\rm *,sph}/\upsilon\,{\rm M}_\odot) - C] +B$ } (Fig.~\ref{Fig-3-4}a)\\
E   (34)     &  1.68$\pm$0.17  &  11.45  & 9.00$\pm$0.12 &  0.38 \\
S0  (30)     &  1.65$\pm$0.19  &  10.30  & 8.00$\pm$0.14 &  0.43 \\
S   (25)     &  2.66$\pm$0.67  &  10.14  & 7.39$\pm$0.14 &  0.57 \\
\hline
\multicolumn{5}{c}{$\log(M_{\rm bh}/{\rm M}_\odot) = A[\log(M_{\rm dyn,sph}/{\rm M}_\odot) - C] +B$} (Fig.~\ref{Fig-3-4}c)\\
E   (34)     &  1.58$\pm$0.18  &  11.78  & 8.99$\pm$0.12 &  0.43 \\
S0  (30)     &  1.25$\pm$0.13  &  10.30  & 8.02$\pm$0.15 &  0.37 \\
S   (25)     &  1.55$\pm$0.24  &  10.05  & 7.39$\pm$0.16 &  0.56 \\
\hline
\multicolumn{5}{c}{$\log(M_{\rm bh}/{\rm M}_\odot) = A[\log(\sigma/{\rm km\,s}^{-1})-C] +B$} (Fig.~\ref{Fig-3-5}) \\
E   (34)     &  7.27$\pm$0.91  &  2.41 & 8.99$\pm$0.12 &  0.43 \\
S0  (30)     &  5.68$\pm$0.60  &  2.24 & 8.03$\pm$0.15 &  0.45 \\ 
S   (25)     &  6.19$\pm$1.11  &  2.13 & 7.39$\pm$0.16 &  0.68 \\
All (89)     &  5.90$\pm$0.33  &  2.27 & 8.22$\pm$0.10 &  0.52 \\ 
\hline
\multicolumn{5}{c}{$\log(M_{\rm *,sph}/ \upsilon {\rm M}_\odot) = A[\log(R_{\rm e,sph,eq}/{\rm kpc})-C] +B$} (Fig.~\ref{Fig-3-6}) \\
E   (34)     &  1.51$\pm$0.17  &     0.89  & 11.44$\pm$0.08 &   0.24 \\
S0  (30)     &  1.22$\pm$0.13  &  $-$0.24  & 10.32$\pm$0.10 &   0.28 \\
S   (25)     &  1.03$\pm$0.22  &  $-$0.28  & 10.13$\pm$0.08 &   0.31 \\ 
All (89)     &  1.10$\pm$0.04  &     0.20  & 10.72$\pm$0.08 &   0.28 \\
\hline
\end{tabular}

For the Bayesian model fitting, 
a fractional error of 13, 20 and 25 per cent was assigned to the velocity
dispersions, radii and dynamical mass, while the individual errors on the spheroid
stellar mass and black hole mass, tabulated in \citet{Graham:Sahu:22a}, were
used. 
Note: the Bayesian analysis was not designed to minimise the scatter in the
vertical direction, reported here as $\Delta_{\rm rms}$.  
The lower-case $\Upsilon$ term, $\upsilon$, allows for conversions between
adopted stellar mass-to-light ratios.  For this study, it is equal to 1.0.
Within the statistical analyses, the Monte Carlo method results in slopes and
intercepts that differ by a few hundredths from run to run. This can be seen
in some of the slightly different intercepts above.  This variation is, however,
well contained within the tabulated parameter uncertainties.

\end{table}

Given the trend in Fig.~\ref{Fig-3-3}, the separation between the E galaxies
and both the S and S0 galaxies is greater in the $M_{\rm bh}$-$M_{\rm
  dyn,sph}$ diagram (Fig.~\ref{Fig-3-4}, right-hand panel) than it is in the
$M_{\rm bh}$-$M_{\rm *,sph}$ diagram (Fig.~\ref{Fig-3-4}, left-hand panel), 
with increased separation of roughly 
$\log(2.3)=0.36$ dex in the $\log M_{\rm dyn,sph}$ direction.

Fig.~\ref{Fig-3-6} reveals that while there is a tight $M_{\rm
  *,sph}$-$R_{\rm e,sph}$ relation for spheroids, there is a tendency for the
E and ES,e galaxies to have larger sizes than the bulges of S, S0 and ES,b
galaxies at a given stellar mass.  
This offset has previously been shown in \citet[][their Fig.~13]{2008MNRAS.388.1708G}
and \citet[][his Fig.~13]{2009MNRAS.393.1531G}. 
This explains why the offset between 
these populations in the $M_{\rm bh}$-$R_{\rm e,sph}$ diagram 
(Fig.~\ref{Fig-3-4}, left-hand panel) is slightly
greater than it is in the $M_{\rm bh}$-$M_{\rm *,sph}$ diagram
(Fig.~\ref{Fig-3-4}, middle panel).  Had all
systems exactly followed the (steeper than linear) $M_{\rm *,sph} \propto
R_{\rm e,sph}^{1.12}$ relation (see Table~\ref{Table-IP13}), one would
have expected the size of the offsets to be reversed, i.e., slightly smaller in the
$M_{\rm bh}$-$R_{\rm e,sph}$ diagram.

The inclusion of BCGs with elevated sizes, due to the inclusion of intracluster
light (ICL), will lead to a reduction of slope in the $M_{\rm *,sph}$-$R_{\rm
  e,sph}$ diagram for the E galaxies.  As noted in \citet{Graham:Sahu:22b},
while the sample used here contains BCGs, these BCGs are not expected to have
elevated sizes due to the ICL.  For reference, at $M_{\rm *,sph} >
3\times10^{10}$ M$_\odot$, \citet{2021MNRAS.504.3058M} report a slope in the
$M_{\rm *,sph}$-$R_{\rm e,sph}$ diagram of 1.15$\pm$0.05, agreeing 
with the trend seen here for all spheroids.  \citet[][their
  Fig.~8d]{2001A&A...368...16M} report a slope of 1.19 between the near-IR
luminosity and half-light radii of bulges in bright spiral galaxies.

However, in passing it is noted that \citet{2010MNRAS.405.1089L} and \citet[][their
  Eq.~3]{2007ApJ...665.1104B} report, respectively, slopes of 1.6 and 2.6 for
S0 bulges.  Furthermore, \citet{2009MNRAS.393.1531G} analysed SDSS data and
reported notably different slopes in the $M_{\rm *,sph}$-$R_{\rm e,sph}$
diagram of 5, 3.33, and 2.63 for what they considered pseudobulges,
classical bulges, and elliptical galaxies, respectively.  In contrast, the
current investigation finds slopes of roughly 1 to 1.5 for the S, S0 and E
samples in the mass range shown.  Although, there are no spheroids with masses
below $\sim$2$\times10^9$ M$_\odot$ in these samples, and they appear to introduce some
curvature into, and steepening of, the relation for bulges, as seen in 
\citet[][their Fig.~8a]{2010MNRAS.405.1089L}, 
\citet[][his Fig.~17 and 18]{2019PASA...36...35G}, based on data in
\citet[][their Fig.~12]{2008MNRAS.388.1708G}, 
\citet[][their Fig.~8]{2021MNRAS.504.3058M}, and thus \citet{2016MNRAS.462.1470L}.
Nonetheless, the analysis of the present sample's SST images and the analysis
of SDSS images by \citet[][their
  Fig.~16]{2022MNRAS.514.3410H} --- obtained from multicomponent
decompositions which account for potentially biasing structures such as bars,
inner discs, nuclear star clusters or depleted cores --- find sizes at $M_{\rm *,sph}\sim
2\times 10^9$ M$_\odot$ which are 2 to 3 times smaller than suggested by the
above works.  This is explored further in \citet{HGS2022}.

\subsection{Quantifying Dry Mergers}

Returning to the $M_{\rm bh}$-$\sigma$ diagram,  the impact
of dry E-building mergers is quantified here. 

Considering an equal mass E$+$E merger in which 
$M_{\rm bh}$ doubles and thus $\log M_{\rm bh}$ 
increases by $\sim$0.3 dex, the jump in $\log R_{\rm e,sph}$ in the 
$M_{\rm bh}$-$R_{\rm e,sph}$ diagram ($0.30/2.50=0.12$)
and the jump in $\log M_{\rm dyn,sph}$ in the 
$M_{\rm bh}$-$M_{\rm dyn,sph}$ diagram
($0.3/1.58=0.19$)  is associated with an expected jump of
0.035 dex in $\log\,\sigma$.  This stems from Equation~\ref{Eq_dyn}, from
which one has $\delta \log M_{\rm dyn,sph} = 2\delta \log \sigma + 
\delta \log R_{\rm e,sph}$. 
It translates to movement along a slope of $\sim$8.6 
($=0.3/0.035$) in the $M_{\rm bh}$-$\sigma$ diagram.  That is, a steepening is
predicted at high masses due to E$+$E mergers. Mergers between ES galaxies 
and between ES and E galaxies are expected to yield a similar result. 
Such mergers will contribute to the 
steeper $M_{\rm bh}$-$\sigma$ relation observed by \citet{2018ApJ...852..131B} for 
BCGs and by \citet{2019ApJ...887...10S} for core-S\'ersic galaxies and those
with high velocity dispersions $>$270 km s$^{-1}$. 
\citep{2013ApJ...764..151G}.  
This trend can be compared with idealised simulations and analytic studies of
dry and gas-poor mergers
between E galaxies on parabolic orbits, in which the velocity dispersion does
not increase by much \citep[e.g.,][see their~Fig.~3]{2001ApJ...552L..13C,
  2003MNRAS.342..501N, 2013ApJ...768...29V}. 

Another way to tackle this is to consider an E$+$E galaxy merger in
which the dynamical, spheroid and black hole mass double.  From the E galaxy
$M_{\rm *,sph}$-$R_{\rm e,sph}$ 
relation (Fig.~\ref{Fig-3-6}), one has $\log M_{\rm *,sph} \propto
1.51\log R_{\rm e,sph}$ (see Table~\ref{Table-IP13}, 
and thus a 0.30 dex increase in stellar mass corresponds 
to a 0.20 (=0.30/1.51) dex increase in spheroid size.  
The virial relation $\log M_{\rm dyn,sph} \propto \log R_{\rm e,sph} +
2\log\sigma$ informs us that a 0.30 dex increase in the dynamical mass,
coupled with a 0.20 dex increase in $R_{\rm e,sph}$, requires a 0.05 dex
increase in $\sigma$.  Now, given the associated 0.3 dex increase in $M_{\rm
  bh}$, this leads to movement along a slope of 6 (=0.3/0.05) in the $M_{\rm
  bh}$-$\sigma$ diagram.  This approach works when $R_{\rm e,sph}$ and
$\sigma$ are suitable for use in the virial theorem.  To illustrate the need
for caution, \citet{2019PASA...36...35G} and \citet{2020ApJ...903...97S}
reported on how the stellar mass-size relation for ETGs and spheroids changed
for sizes defined by radii enclosing different fractions of light than the
canonical 50 per cent value associated with the effective half-light radii,
$R_{\rm e}$.  These changes affects both typical dynamical mass estimates
(based on $\sigma^2R$) and implications for baryon-to-(dark matter) ratios. 
This is left for now, but further discussion is provided in Appendix~\ref{Sec_Apdx}.

Next, consider equal mass S0$+$S0 galaxy mergers taking systems from
$\log(M_{\rm bh}/{\rm M}_\odot)$ equals 9.0 to 9.3 dex and creating an E
galaxy; that is, moving from the S0 galaxy to the E galaxy  $M_{\rm bh}$-$M_{\rm
  dyn}$ and $M_{\rm bh}$-$R_{\rm e,sph}$ relations. 
Following the above procedure, one can calculate changes of 0.0611 dex in 
$\log\,\sigma$ and a shift along a line with a slope of $\sim$4.9. Such
mergers, therefore, largely move systems along the $M_{\rm bh}$-$\sigma$
relation.  This movement explains why the $M_{\rm bh}$-$\sigma$
relation has less horizontal scatter than the 
$M_{\rm bh}$-$M_{\rm *,sph}$ and $M_{\rm bh}$-$R_{\rm e,sph}$ relations in
which the E galaxies are clearly offset from the S0 galaxies due to their
lower $M_{\rm bh}/M_{\rm *,sph}$ ratios which arose when mergers contributed
disc stars, as detailed in \citet{Graham:Sahu:22a}. 

Another observation leading credence to the merger-built E galaxies is made in
passing.  The stellar density, $\rho_{\rm *,sph}$, within a sphere of radius
$R$ is proportional to $M_{\rm *,sph}/R_{\rm sph}^3$.  Using $M_{\rm *,sph}
\propto R_{\rm e,sph}^{1.10}$ (as an approximate trend for all spheroids, see
Table~\ref{Table-IP13}), one has that $\rho_{\rm *,sph} \propto R_{\rm
  e,sph}^{1.90}$.  This relation is why the density of smaller bulges in disc
galaxies is greater than that of larger elliptical galaxies, as shown in
\citet{2013pss6.book...91G}.  In terms of a dry S0$+$S0 merger building an E
galaxy, one can think of the pre-merger (initially disc) stars at larger radii
dispersing to create an elliptical galaxy with its light profile having a
shallow quasi-exponential tail.  This likely also explains the red-to-blue
colour gradient observed in some E galaxies
\citep[e.g.,][]{1976ApJ...204..684S, 1990JKAS...23...43K} if comprised of
former disc stars at large radii and former bulge stars at small radii.  It
also helps explain why such mergers produce the large jump in $R_{\rm e,sph}$,
along with the jump in $M_{\rm *,sph}$, because $R_{\rm e,gal}$ is 
notably larger than $R_{\rm e,sph}$ for most S0 galaxies.

Finally, consider a dry, equal mass S0$+$S0 galaxy merger taking two
lower-mass S0 galaxies
with $\log(M_{\rm bh}/{\rm M}_\odot)$ equal to 7.7 to one E galaxy with
$\log(M_{\rm bh}/{\rm M}_\odot) = 8.0$.  This is accompanied by a shift in
$\log\,\sigma$ of 0.10 dex, corresponding to movement along a line with a slope
of 3.0 in the $M_{\rm bh}$-$\sigma$ diagram.  
This current S0$+$S0 example merger corresponds to a shift in spheroid mass of
0.74 dex (see Fig.~\ref{Fig-3-4}), which could be achieved\footnote{The
  initial S0 galaxy $B/T$ ratios are simply twice their spheroid's stellar
  mass divided by the total stellar mass of the E (merger remnant) galaxy.}
by the dry merger of two similar S0 galaxies with bulge-to-total ratios of
0.37 prior to merging.\footnote{If the merger remnant is an S0 or ES galaxy
  with a disc component, the original $B/T$ ratio would have been smaller than
  0.37.}
This merger is associated with a slight shift away (to higher velocity
dispersions) from the mean $M_{\rm bh}$-$\sigma$ relation at mid-(black hole 
masses).  This can be seen through the red line for E galaxies residing on the
right-hand side of the cyan line for S0 galaxies at $M_{\rm bh}\sim10^8$
M$_\odot$ in the $M_{\rm bh}$-$\sigma$ diagram (Fig.~\ref{Fig-3-5}).  Coupled
with the smaller shift (in $\log \sigma$) at higher black hole masses --- due
to E$+$E mergers --- this will contribute to, if not explain, the
observed steepening in the $M_{\rm bh}$-$\sigma$ diagram, seen in
Fig.~\ref{Fig-3-5} and reported by \citet{2018ApJ...852..131B} and
\citet{2019ApJ...887...10S}.  Curiously, it is noted that at black hole masses
below $\sim$$10^8$ M$_\odot$ and $\log(\sigma\, {\rm km}\, {\rm s}^{-1})
\lesssim 2.2$ (or $\sigma \lesssim 160$ km s$^{-1}$), the $M_{\rm
  bh}$-$\sigma$ relation appears less well defined, although this may be a
consequence of the small data range and roughly $\pm$0.5 dex ($\pm$1 sigma)
scatter in the $\log M_{\rm bh}$ direction.
%

\subsubsection{Wet mergers}

The notion that E galaxies might be built from the merger of two (bulgeless or
near bulgeless) disc galaxies \citep[e.g.,][]{1977A&A....54..121V,
  1977egsp.conf..401T} is problematic.  The early simulations which gave rise
to this idea contained no gas.  Such mergers would yield a large jump in
spheroid mass which may not be matched by the even greater jump in black hole
mass required to maintain the steeper-than-linear $M_{\rm bh}$-$M_{\rm *,sph}$
trend.  For example, consider the hypothetical creation of an E galaxy from
the merger of two equal spiral galaxies having $B/T=0.1$, 
$M_{\rm *,sph}= 0.5\times 10^{10}$ M$_\odot$, and 
$M_{\rm bh}= 0.5\times 10^7$ M$_\odot$. 
Prior to any black hole fueling or star
formation, this hypothetical system will have 
$M_{\rm *,sph}= 10^{11}$ M$_\odot$, and 
$M_{\rm bh}= 10^8$ M$_\odot$.  However, E galaxies with these properties are
not known (see the middle panel of Figure~\ref{Fig-3-4}). 
If there was, however, sufficient quasar activity to drive up the
black hole mass to meet with the spheroid stellar mass on the observed $M_{\rm
  bh}$-$M_{\rm *,sph}$ relation, then this scenario does not represent the
much-acclaimed symbioses in which the black hole regulates the spheroid's star
formation to establish the black hole scaling relations.  

Even if the tidally-engaged but not yet wed galaxies experienced bar
formation, which led to cold gas inflow that fuelled a bulge-building central
starburst \citep[e.g.,][]{1996ApJ...471..115B}, many of the stars in the
proposed E galaxy merger remnant would still have come from the discs of the
two single galaxies.  Once again, any quasar-fuelled growth of the merging
black holes has not regulated the mass of these pre-existing disc stars, which
will likely dominate the final stellar mass budget. This tension is alleviated
if spiral galaxy collisions do not build E galaxies but instead build disc
galaxies with less massive spheroids.  This gels with binary disc simulations
in which the system's net orbital angular momentum is not nulled by their
merger \citep[e.g.,][]{2006MNRAS.372..839N}, and with the need for dissipation
in the formation of some ETGs, that is, some S0 galaxies \citep[see][]{2001ApJ...552L..13C}.  As discussed in
\citet{Graham:Sahu:22b}, this scenario also meshes well with the disc 
structures observed in the remnants of (cold gas)-rich mergers, and it can
readily accommodate the $M_{\rm bh}/M_{\rm *,sph}$ ratios in these remnants.

While the current sample size of five such (cold gas)-rich merger products is
small, their alignment with (i) the E galaxy's $M_{\rm bh}$-$M_{\rm *,sph}$
relation but also (ii) the spiral galaxy's $M_{\rm bh}$-$R_{\rm e,sph}$
relation, requires consideration beyond the present analysis of how dry
mergers influence the $M_{\rm bh}$-$\sigma$ diagram.  Potential evolutionary
pathways from the S to the S0 sequence are presented in
\citet{1998A&A...331L...1S} for the $M_{\rm bh}$-$\sigma$ diagram, and in
\citet{Graham:Sahu:22a} for the $M_{\rm bh}$-$M_{\rm *,sph}$ diagram, while
Graham (2022, in preparation) presents a more complete picture in which
S0 galaxies can also evolve into S galaxies. 

While discussing mergers involving spiral galaxies, it is noted that some
spiral galaxies may acquire (their initial) and/or grow their central massive
black hole via the accretion of lower-mass dwarf galaxies.  For example,
Nikhuli, the suspected remains of a dwarf galaxy's nuclear star cluster, may
be delivering an intermediate-mass black hole (IMBH) to NGC~4424
\citep{2021ApJ...923..146G}.  
Chandra X-ray Observatory imaging of 35 spiral
galaxies expected to harbour an IMBH has recently revealed central X-ray
point sources in 14 of them \citep{2021ApJ...923..246G}, possibly signalling
the suspected abundance of IMBHs \citep{2017ApJ...839L..13S}.  Meanwhile, a similar
campaign which looked at dwarf ETGs found just three candidates from a sample of 30
\citep{2008ApJ...680..154G, 2019MNRAS.484..794G}, perhaps due to the reduced
fuel supply, and thus quieter black holes, in these relatively (cold gas)-poor systems.


Finally, it is noted that increases in black hole mass due to the
collision of massive black holes does not void the argument of
\citet{1982MNRAS.200..115S}, which makes a case for quasar-fuelled black hole
growth explaining the bulk of the mass locked up in black holes today
\citep[e.g.,][]{2004MNRAS.354.1020S}.  What is presented here is a case for an addendum
to the Soltan argument, or at least additional knowledge enabling a fuller 
understanding, in which the black hole mass function undergoes evolution not
captured by the Soltan scenario.  Dry mergers neither increase the total amount
of mass locked up in black holes nor do they negate past quasar activity, but
they do require it be scaled back in the most massive,
merger-built, black holes.  As discussed in \citet{Graham:Sahu:22a}, quasars
and Seyferts tend to occur in disc galaxies rather than massive E galaxies,
which have hot/radio mode activity rather than cold/quasar mode activity.


\subsection{A consistency check}


Following \citet{2012ApJ...746..113G}, a consistency check is performed as
to whether the $M_{\rm bh}$-$M_{\rm *,sph}$ and $M_{\rm bh}$-$\sigma$
relations combine to give $M_{\rm *,sph}$-$\sigma$ relations 
consistent with expectations.  
\citet[][their Table~5]{2019ApJ...887...10S} have reported slopes 
of 2.10$\pm$0.41, 2.97$\pm$0.43, and 5.16$\pm$0.53
for the $M_{\rm *,sph}$-$\sigma$ relations 
for late-type galaxies (LTGs), S\'ersic ETGs, and core-S\'ersic galaxies,
respectively.\footnote{These slopes are based on the use of $M_*/L_{*,3.6}=0.6$.} 
These slopes are broadly in accord with 
samples of low-to-intermediate ETGs \citep{1981ApJ...251L...1T} which used
galaxy rather than spheroid magnitudes, 
and samples of massive ETGs \citep{1980AJ.....85..801S} likely dominated by
elliptical galaxies.  The slopes for the S, S0 and E galaxies 
in the $M_{\rm *,sph}$-$\sigma$ diagram (Fig.~\ref{Fig-3-1}) 
are 2.49$\pm$0.56, 3.48$\pm$0.47, and 4.24$\pm$0.54. 
Combining these slopes with those from the  
$M_{\rm bh}$-$M_{\rm *,sph}$ relations in Table~\ref{Table-IP13} yields 
expected slopes for the $M_{\rm bh}$-$\sigma$ relation 
of 
6.62 ($=2.49\times2.66$), 
5.74 ($=3.48\times1.65$), and 
7.12 ($=4.24\times1.68$).  
This is in accord with the above explanation surrounding the
merger-induced transformation of galaxies.


\section{Summary and concluding remarks}
\label{Sec_Summary}

It has been explored how the largest black holes came to be, with vital input from
published galaxy morphology analyses which separated bulges from discs (and other
galaxy components),  
yielding spheroid sizes and masses for use in  Clausius' 
far-reaching mechanical theorem applicable to heat \citep{1870AnP...217..124C}. 
Better known as the `virial theorem', it is routinely applied 
to pressure-supported stellar systems. 
%
%
Comparing the dynamical-to-stellar masses of the spheroids has enabled one to 
understand the changes in the velocity dispersion arising from S0 mergers and
the emergence of the E galaxies, with their lower $M_{\rm bh}/M_{\rm *,sph}$ ratios
at a given spheroid mass.  
While such dry galaxy mergers can yield a substantial jump in the $M_{\rm
 bh}$-$M_{\rm *,sph}$ diagram, the 
virial theorem was used to explain why there is not a substantial 
jump in the $M_{\rm bh}$--(velocity dispersion, $\sigma$) diagram.
It is because the jump in $M_{\rm *,sph}$ and $M_{\rm dyn,sph}$ is largely matched by
a jump in $R_{\rm e,sph}$, leaving little room for $\sigma$ to change.

It was noted that the frequency of late-type spiral galaxy mergers to
produce an E galaxy, as opposed to a bigger disc galaxy, appears low because
the small bulge-to-total stellar mass ratios, $B/T$, of late-type spiral
galaxies would be associated with a substantial horizontal jump in the $M_{\rm
  bh}$-$M_{\rm *,sph}$ diagram.  This would need to be matched by dramatic
AGN fuelling. For instance, late-type spiral galaxies with $B/T\approx 0.1$
would invoke a jump in spheroid mass by a factor of 20 if such galaxies are to
collide and build an E galaxy.  As seen in the middle panel of
Fig.~\ref{Fig-3-4}, this will require the central black hole to increase its
mass by two orders of magnitude if the system is to land on the $M_{\rm
  bh}$-$M_{\rm *,sph}$ relation for E galaxies.  This may be feasible
\citep[][their Fig.~1]{2001ApJ...555..719C, 2022MNRAS.511.5756T}, although it remains challenging
to erase the orbital angular momentum of the disc-dominated progenitors,
unless the orbital configuration of the collision is finely tuned.  
More likely is that the cold gas in these systems will fuel the AGN and also lead
to significant star formation, building something akin to the S0 galaxy
NGC~5128 --- one of the five pink hexagons in the figures.  

The ES,e type of ellicular galaxy may be an evolutionary stepping stone between S0
and E galaxies, with subsequent ES,e galaxy mergers building yet more massive E
galaxies. 
Combined with E+E galaxy mergers producing some of the brightest cluster galaxies,
the march of galaxies can be tracked in multiple black hole scaling diagrams.  
It is apparent that the black hole scaling relations for E/ES,e galaxies are a result
of mergers, with the initially-disc-stars contributing to the mass, size, 
and velocity dispersion of the remnant spheroid.  This then forms an important
complement to AGN feedback models which establish the hole scaling
relations in the non-elliptical, i.e., disc, galaxies, such as the S0 galaxy 
 $M_{\rm bh}$-$M_{\rm *,sph}$ relation with its order of magnitude higher
normalisation than the E galaxy $M_{\rm bh}$-$M_{\rm *,sph}$ relation. 
It also offers an explanation for E galaxy $M_{\rm bh}$-$\sigma$ relation
with its steep slope $\sim$7$\pm$1. 

AGN feedback may establish the $M_{\rm bh}$-$\sigma$ relation for the S and
S0 galaxies, but mergers are also a part of the story, from the possible
delivery of IMBHs into spiral galaxies to the creation of the biggest black
holes in massive E galaxies.  This work is not claiming that mergers can explain
the full distribution in the $M_{\rm bh}$-$\sigma$ diagram, but it explains 
why dry mergers do not create large `punctuated equilibrium'-sized
jumps in the $M_{\rm bh}$-$\sigma$ diagram. 
It was revealed
how these dry mergers move systems somewhat parallel to, or at least not too far
from, the $M_{\rm bh}$-$\sigma$ relation, thereby maintaining, if not
creating, a low level of scatter, at least in the upper half of this diagram
where ES,e and E galaxies reside. This work additionally explains
and quantifies the steepening of the $M_{\rm bh}$-$\sigma$ relation due to
the production of Es,e and E galaxies.

The revelation, and refinement, of (galaxy morphology)-dependent black hole
scaling relations over the past decade has consequences for many 
astrophysical phenomenon related to massive black holes.  One of these is 
the generation of long-wavelength gravitational radiation associated with 
massive black holes 
\citep{2010CQGra..27h4013H, 2019A&ARv..27....5B}. Although it has long been
recognised that E galaxies form from mergers, there has been some latency in
establishing the E galaxy black hole scaling relations. There are, of course, 
additional merger remnants than just E/ES,e galaxies in which black holes are expected to coalesce, such
as the core-S\'ersic S0 galaxies and the gas-rich mergers like the 
Centaurus galaxy.  \citet[their Eq.~1][]{Graham:Sahu:22b} provide a general,
and hopefully useful, single 
$M_{\rm bh}$-$M_{\rm *,gal}$ relation for all of these merger-built systems. The 
logarithmic slope of the relation is $\sim$2, dramatically different from the
slope of $\sim$1 which has often been used in the past. 
To give just one example, \citet{2012A&A...542A.102M} revealed
how a steeper slope of $\sim$2, rather than $\sim$1, for the 
$M_{\rm bh}$-$M_{\rm *,sph}$ relation yields an order of magnitude reduction to the
expected detection rate of extreme mass ratio inspiral (EMRI) events by the
Laser Interferometer Space Antenna \citep[LISA:][]{1997CQGra..14.1399D}.

Application of updated (galaxy morphology)-dependent black hole scaling relations may prove
beneficial for expectations from, and design of, some gravitational wave
detectors, such as the 
Einstein Telescope \citep[ET:][]{2010CQGra..27s4002P, 2011GReGr..43..485G,
  2011PhRvD..83d4020H}, 
the Cosmic Explorer \citep[CE:][]{2019BAAS...51g..35R, 2021arXiv210909882E}, 
the Deci-Hertz Interferometer Gravitational wave Observatory
\citep[{\it DECIGO}:][]{2011CQGra..28i4011K, 2021Galax...9...14I}, and 
TianQin \citep{2016CQGra..33c5010L, 2021PTEP.2021eA107M}. 
It should also aid with expectations for, and assist with implications of,
data from pulsar timing array (PTA) projects such as that at the Parkes radio
telescope 
\citep[PPTA:][]{2013PASA...30...17M, 2021ApJ...917L..19G}, 
the European PTA
\citep[EPTA:][]{2006ChJAS...6b.298S, 2021MNRAS.508.4970C}, 
and other promising ventures \citep[e.g.,][]{2010CQGra..27h4013H, 2018JApA...39...51J,
  2020PASA...37...28B}. 
Excitingly, 
the North American Nanohertz Observatory for Gravitational Waves 
(NANOGrav) recently reported a detection \citep{2020ApJ...905L..34A} 
which might be consistent with primordial solar-mass and planet-mass black
holes \citep{2021PhLB..81336040K, 2022SCPMA..6530411D}.  PTAs such as this
one 
are not sensitive to the higher frequency gravitational radiation associated with 
primordial black holes (PBHs) less massive than $\sim$10$^{-10}$~M$_\odot$, which
is the regime where a population of PBHs might explain all dark matter. 
Nonetheless, 
this has understandably re-sparked interest in the possibility that dark matter may be PBHs 
\citep[e.g.,][]{1994PhRvD..50.7173I, 2002PhLB..535...11B, 2021PhRvL.126d1303D, 2022arXiv220814279G}.


\section*{Acknowledgements}

AWG is grateful for past discussions with Anita Pappas, David Brown, Mike
Russell, and Joe Silk. 
Part of this research was conducted within the Australian Research Council's
  Centre of Excellence for Gravitational Wave Discovery (OzGrav) through
  project number CE170100004.
This work has used the NASA/IPAC Infrared Science Archive (IRSA)
and the NASA/IPAC Extragalactic Database (NED), 
funded by NASA and operated by the California Institute of Technology.
This research has also used NASA's Astrophysics Data System Bibliographic
Services and the \textsc{HyperLeda} database
(\url{http://leda.univ-lyon1.fr}). 
This study used the {\sc Rstan} package available at \url{https://mc-stan.org/}.

\section{Data Availability}

The imaging data underlying this article is available in the NASA/IPAC Infrared
Science Archive, while the kinematic data was sourced from HyperLeda.
The derived spheroid masses and sizes are
tabulated in \citet{Graham:Sahu:22a}, while the velocity dispersions shown in
Fig.~\ref{Fig-3-5} are conveniently listed in \citet{2019ApJ...887...10S}.

\bibliographystyle{mnras}
\bibliography{Paper-BH-sigma}{}

\appendix

\section{Virial coefficients}
\label{Sec_Apdx}

\begin{figure*}
\begin{center}
\includegraphics[trim=0.0cm 0cm 0.0cm 0cm, width=1.0\textwidth, angle=0]{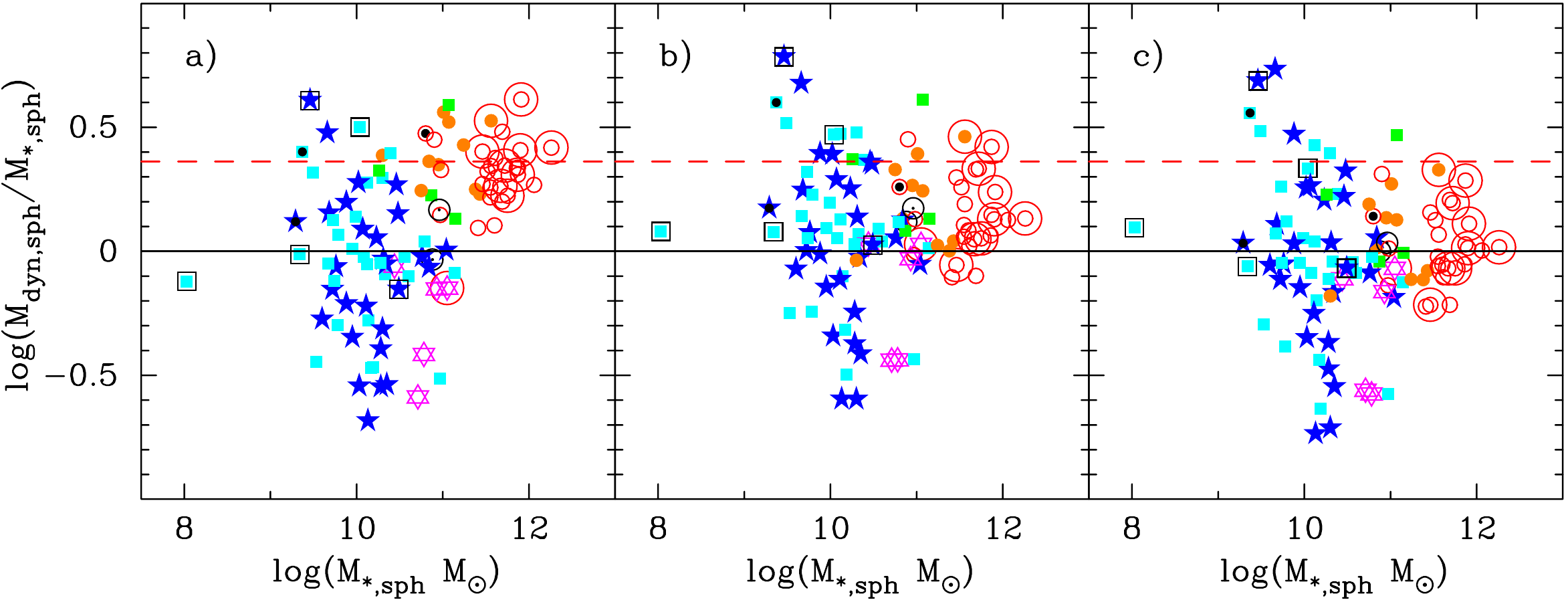}
\caption{Panel a) variant of Fig.~\ref{Fig-3-3}.  Symbols have the same
  meaning. 
Panel b) shows a modification of panel a), now using 
$M_{\rm dyn,sph} = K_v(n) \sigma^2 R_{\rm e,sph}$, where $K_v(n)$ is
 given by equation~\ref{Eq_Kv}.  
Panel c) also shows a modification of panel a), now using 
$M_{\rm dyn,sph} = 5\times(5 \sigma^2 R_{\rm 0.1,sph}$), where $R_{\rm 0.1,sph}$ is 
the radius enclosing the central 10 per cent of the spheroid light, i.e.,
one-fifth of that enclosed by $R_{\rm e,sph}$. As in Fig.~\ref{Fig-3-3}, 
the arbitrary dashed line is at $M_{\rm dyn,sph}/M_{\rm *,sph} = 2.3$. 
}
\label{Fig-3-7}
\end{center}
\end{figure*}

In the 1990s, researchers started to explore the implications of broken structural homology,
i.e., the observation that bulges and elliptical galaxies are not described
by \citet{1948AnAp...11..247D} 
`$R^{1/4}$ law' but instead by \citet{1963BAAA....6...41S} `$R^{1/n}$
model'.  The variety of stellar density profiles, quantified by the 
S\'ersic index, $n$, implied a variety of velocity 
dispersion profiles \citep{1991A&A...249...99C} and thus aperture velocity
dispersion profiles (\citet[][their Fig.~7]{1997A&A...321..724C};
\citet[][their Figures~7--8]{1997MNRAS.287..221G}).  These are required in order to uphold these
pressure-supported systems. 
The range of profiles means that the measured central velocity dispersion,
$\sigma$, relative to the desired virial velocity dispersion 
depends not only on the
size of the central aperture but also on the spheroid's S\'ersic index.
To avoid the practical difficulties with reading from a figure, 
\citet[][see their Figures~9--11]{1997A&A...321..111P} also 
provided a useful table 
accounting for the different profile shapes when an aperture velocity
dispersion is measured within 0.1
$R_{\rm e,sph}$.  Their tabulated $S_D(n)$ term is the virial
coefficient\footnote{This should not be confused with the virial factor, $f$,
used to convert AGN virial masses into black masses 
\citep[e.g.,][]{1993PASP..105..247P, 1998ApJ...505L..83L}.  The value of $f$
is often derived using the $M_{\rm bh}$-$\sigma$ relation coupled with
reverberation mapping data \citep[e.g.,][]{2004ApJ...615..645O,
  2011MNRAS.412.2211G, 2019MNRAS.488.1519Y}.}, when using 
$\sigma=\sigma_{\rm ap}(R<0.1R_{\rm e})$, for calculating the virial mass via 
\begin{equation}
M_{\rm virial} = S_D(n) \sigma^2 R_{\rm e} / G.
\end{equation}
\citet[][their Fig.~D.1]{2002A&A...386..149B} refer to
$S_D(n)$ as $K_V(n)$, and for velocity dispersions measured in central 
apertures of $R_{\rm e}/8$ or $R_{\rm e}/10$,
their Equation~11 parameterised these tabulated virial coefficient as 
\begin{equation}
K_V(n) \approx \frac{73.32}{10.465+(n-0.94)^2} + 0.954.
\label{Eq_Kv}
\end{equation}  

In Fig.~\ref{Fig-3-7}b, Equation~\ref{Eq_Kv} was used to
derive a virial mass for the spheroids.  This mass differs from panel a), in which
a constant virial coefficient of 5 was used (see Equation~\ref{Eq_dyn}).  
The difference between the two panels in Fig.~\ref{Fig-3-7} is 
similar to that seen in \citet[][their Figures~12--13]{2008MNRAS.389.1924F}.
Pursuing this further is beyond the scope of the present investigation into 
the $M_{\rm bh}$-$\sigma$ relations, for which a constant
virial coefficient suffices.  However, the use of the (S\'ersic $n$)-dependent $K_v(n)$
term in Fig.~\ref{Fig-3-7} serves as a reminder of why one cannot yet infer
trends regarding potential fractions of dark matter.  This point is also illustrated
through the use of a different scale radius.  As detailed in
\citet{2019PASA...36...35G}, the use of the half-light radius, $R_{\rm e}$,
has always been arbitrary, without any physical significance beyond containing
50 per cent of the light.  

Moreover, its use has led to the misunderstanding of galaxy connections.
Here, the use of scale radii containing 10 per cent of the spheroid light is
briefly explored.  To help offset this reduction from the 50 per cent
half-light radius, the coefficient in Equation~\ref{Eq_dyn} is increased by a
factor of 5.

In Fig.~\ref{Fig-3-7}c, a variant of 
Equation~\ref{Eq_dyn} has been used, such that 
\begin{equation}
M_{\rm dyn} = 5\times \left(5 \sigma^2 R_{\rm 0.1} / G \right).
\end{equation}
Here, $R_{\rm 0.1}$ is the projected radius on the sky enclosing the central 10 per cent
of the spheroid light.  The equation for the transformation between $R_{\rm
  e}\equiv R_{0.5}$ and $R_{\rm 0.1}$ can be found in \citet[][his
  Equation~22]{2019PASA...36...35G}. This size transformation equation is a function of the
S\'ersic index, $n$.  

There are several additional factors regarding the virial coefficient.  For
example, while the E galaxies may have an average $R_{\rm e}$ of around 6~kpc,
and thus the 0.595~kpc aperture velocity dispersions
\citep[see][]{1998A&AS..132..255G} from HyperLeda are, on average, measured at
0.1 $R_{\rm e}$, there is a range of $R_{\rm e}$ from $\sim$2 to 20~kpc for
the E galaxies used here.  Moreover, the bulges of the disc galaxies have
sizes down to $\sim$100~pc.

In the 1990s, it also became apparent that pressure-supported elliptical
galaxies were not as common as previously thought and that discs in ETGs were
much more abundant than had been realised \citep{1990ESOC...35..231C,
  1990ApJ...362...52R, 1998A&AS..133..325G, 2003AJ....126.1787G}. 
Unless kinematic decompositions are performed on the galaxy spectra, 
the dynamics of the 
disc will contribute to the measured velocity dispersion when the bulge is
small. The use of larger apertures, such as `effective
apertures', will increasingly bias the
slope of the $M_{\rm bh}$-$\sigma$ relation at the low-mass end where the
galaxies' discs dominate over their bulges.  That is to
say, the `effective velocity dispersion', $\sigma_{\rm e}$, within the {\em galaxy}
effective half-light radius of S0 galaxies, also known as `fast rotators',
is not applicable for use in the current form of the virial theorem given $R_{\rm e,gal}$ is many
times greater than $R_{\rm e,sph}$.  Basically, $\sigma_{\rm e}$ is partly the disc
velocity dispersion for the disc galaxies, which typically have $R_{\rm
  e,sph}/h_{\rm disc} \approx 0.2$--0.4, or $R_{\rm 
  e,sph}/R_{\rm e,disc} \approx 0.12$--0.24 \citep[e.g.,][]{2008MNRAS.388.1708G}.

Furthermore, \citet{2001AJ....121..820G} presented the `iceberg' model for
spiral galaxies, in which the bulge-to-disc size ratios of spiral galaxies are
somewhat constant, but the surface brightness of the bulge becomes
increasingly faint relative to the inner disc. This reduction in the
prominence of the bulge light in late-type spiral galaxies results in a
greater contribution of disc light to the measured velocity dispersion.  This
contribution will contribute to an observed flattening of the $M_{\rm
  bh}$-$\sigma$ distribution.  With ongoing efforts by many teams to probe
$M_{\rm bh} \lesssim 10^6$~M$_\odot$ at$\sigma \lesssim 100$
km s$^{-1}$ comes a growing need to be mindful of whether $\sigma$ remains
representative of the spheroid.  If the disc dominates the signal,
it may reduce the measured value of $\sigma$.  This would represent
a break from probing the coevolution of spheroids and their black holes, as
the disc velocity dispersion seems unlikely to be regulated by the mass of the
black hole. 

While the spheroidal component of galaxy light is separated during the
multicomponent decompositions of the galaxy images, there is an implicit assumption that the
central velocity dispersion reflects the spheroid light and is not
contaminated by other components, such as the disc.  This assumption is 
increasingly incorrect in lower mass galaxies where the bulges are smaller and
fainter. 
This implicit assumption can be addressed with two-component kinematic models akin to bulge+disc
luminosity models.  
Following \citet{1995A&A...293...20S} and others, \citet{2020MNRAS.495.4638O} 
have shown how the bulge and disc velocity dispersions can be
separated.  This separation shall be explored in future work, mindful that some of the E
galaxies used here (e.g., NGC: 821; 3377, 3607; 4291, and 4621) 
also display major-axis rotation, sometimes reaching 50-100 km 
s$^{-1}$.  These galaxies are likely 
mergers in which not all of the orbital angular momentum of the stars has been 
cancelled, as discussed in \citet{Graham:Sahu:22b}. 

Moreover, the calculated $K_v(n)$ term (Equation~\ref{Eq_Kv}) for
$\sigma_{0.1}$ is based on the S\'ersic $R^{1/n}$ radial distribution of
stellar matter, treating the spheroids as non-rotating spheres with isotropic
dynamical structure.  As such, this term does not account for the influence of
dark matter.  For those wishing to explore this further, it is noted that
modifications for varying levels of orbital anisotropy, i.e.,
radial-to-tangential orbits and dark matter (or radially varying stellar
mass-to-light ratios), were introduced in the pioneering work by
\citet{1996MNRAS.282....1C} and \citet{1997A&A...321..724C} for spherical
S\'ersic $R^{1/n}$ profiles of E galaxies.  Density-potential-(velocity
dispersion) profiles for triaxial core-S\'ersic models can be found in
\citet{2007MNRAS.377..855T}.

\citet{2011MNRAS.412.2211G} discuss additional caveats regarding 
measurements of the central velocity dispersions. One of these pertains to 
dynamically-hot, (dry merger)-built spheroids in which the coalescence of the binary 
black hole has scoured out the central stellar phase space to create spheroids with
core-S\'ersic light profiles \citep{1980Natur.287..307B, 2003AJ....125.2951G,
  2005LRR.....8....8M, 
  2007ApJ...671...53M}.\footnote{Prior to the core-S\'ersic model, galaxies
  with depleted cores were studied by \citet{1966ApJ...143.1002K},
  \citet{1972IAUS...44...87K}, \citet[][and references
    therein]{1997AJ....114.1771F}.}   Removing a galaxy's inner stars 
results in a new dynamical structure with a reduced central
velocity dispersion \citep{2005MNRAS.362..197T}.  This modification will act 
in a sense to further steepen the high-mass end of the $M_{\rm
  bh}$-$\sigma$ relation, where core-S\'ersic galaxies are found
\citep{2019ApJ...887...10S}. This will be explored in future work, from a
theoretical/mathematical perspective and using $N$-body simulations.

\section{Outliers in the $M_{\rm *,sph}$-$\sigma$ diagram}
\label{App_Out}

NGC~4291 was previously flagged by \citet{2019ApJ...887...10S} as a notable
outlier in the $M_{\rm *,sph}$-$\sigma$ diagram.  In
\citet{2016ApJS..222...10S}, NGC~4291 was modelled as an E galaxy.  It is one
of the lowest mass E galaxies in the sample used here and has the smallest
size of the E galaxies in this sample.  The residual (galaxy minus model)
light profile in \citet[][their Fig.~61]{2016ApJS..222...10S} reveals a
`snake-like' pattern suggestive that the single S\'ersic function fit beyond
the depleted core may be
inadequate.\footnote{If NGC~4291 is reclassified from E to S0, then the partially depleted core in NGC~4291
  \citep{2004AJ....127.1917T, 2012ApJ...755..163D} would make it one of three S0
  galaxies in the sample thought to have been built by a dry merger.}  
  The galaxy might, therefore, have a disc component, and,
depending on its (unknown) bulge-to-total stellar mass ratio, it may better
mesh with the $M_{\rm *,sph}$-$\sigma$ relation for S0 galaxies.


Due to the torquing effect that it may have on the linear regressions,
\citet{2019ApJ...876..155S} flagged and excluded NGC~2787. It is the S0 galaxy
with the lowest spheroid mass and the second smallest spheroid size in the S0
galaxy sample.  The galaxy was brought to attention by
\citep{2003ApJ...597..929E} for containing both a classical bulge and a
`pseudobulge', and \citet{2004AJ....127.2641S} note that it has 
 a polar disc plus two nuclear rings which do not reside in the galaxy's main
 disc plane.   The galaxy has most recently been modelled in
\citet{Graham:Sahu:22b}.  Its prominent nuclear disc within the inner couple
of arcseconds \citep{2010MNRAS.407..969L} may have elevated the measured
velocity dispersion relative to the small $R_{\rm e,sph}=140$~pc
classical bulge.\footnote{Based on the image analysis, the `barlens', bar and
  disc do not dominate the light until beyond $\sim$8$\arcsec$.}  However, it
may be that the spheroid, i.e., the classical bulge, mass is low, partly
explaining its offset in the colour-magnitude diagram \citet[][their
  Fig.~A1]{Graham:Sahu:22a}.  Although, including the `barlens' (12.26 mag at
3.6~$\mu$m, AB) with the bulge (11.66 mag at 3.6~$\mu$m, AB) would boost the
current spheroid luminosity by just 58 per cent, or 0.20 dex.

\citet{2020ApJ...903...97S} excluded the S galaxy NGC~4945 because it is a
more than a 2-sigma outlier in their $M_{\rm bh}$--(S\'ersic $n$) diagram,
possibly indicating the S\'ersic index and thus spheroid mass is not correct.
NGC~4945 has the lowest spheroid stellar mass of the S galaxies.  As with
NGC~2787, this galaxy is excluded due to its ability to skew the regressions.
\citet{2020ApJ...903...97S} flagged and excluded a further three outlying
galaxies, which are retained here.  One of these is NGC~3998, seen as
something\footnote{NGC~3998 resides 2.5-sigma (2.5~$\Delta_{\rm rms}$) from
  the $M_{\rm *,sph}$-$\sigma$ relation for 30 S0 galaxies (constructed
  without NGC~2787).}  of an outlier in Fig.~\ref{Fig-3-1} but its
exclusion/inclusion has no significant impact here.
%
%
Another is NGC~5419, which has since been remodelled in
\citet{Graham:Sahu:22b}.  The third galaxy is NGC~3377, also flagged in
\citet[][their Fig.~8]{Graham:Sahu:22a} and mentioned in Fig.~\ref{Fig-3-2}
but otherwise retained here.  These systems are labelled in Fig.~\ref{Fig-3-1}.

\bsp    
\label{lastpage}
\end{document}